\documentstyle[prd,tighten,aps]{revtex}
\begin{document}
\draft

\twocolumn[\hsize\textwidth\columnwidth\hsize\csname
@twocolumnfalse\endcsname
\renewcommand{\theequation}{\thesection . \arabic{equation} }
\title{\bf Generalized De Sitter Space}

\author{Pedro F. Gonz\'{a}lez-D\'{\i}az}
\address{Centro de F\'{\i}sica ``Miguel Catal\'{a}n'',
Instituto de Matem\'{a}ticas y F\'{\i}sica Fundamental,\\ Consejo
Superior de Investigaciones Cient\'{\i}ficas, Serrano 121, 28006
Madrid (SPAIN)}
\date{September 14, 1999}

\maketitle

\begin{abstract}
This paper deals with some two-parameter solutions to the
spherically symmetric, vacuum Einstein equations which, we
argue, are more general than the de Sitter solution. The global
structure of one such spacetimes and its extension to the
multiply connected case have also been investigated. By using a
six-dimensional Minkowskian embedding as its maximal extension,
we check that the thermal properties of the considered solution
in such an embedding space are the same as those derived by the
usual Euclidean method. The stability of the generalized de
Sitter space containing a black hole has been investigated as
well by introducing perturbations of the Ginsparg-Perry type in
first order approximation. It has been obtained that such a
space perdures against the effects of these perturbations.
\end{abstract}


\vskip2pc]

\renewcommand{\theequation}{\arabic{section}.\arabic{equation}}

\section{\bf Introduction}
\setcounter{equation}{0}

In some respects de Sitter space is to cosmology what hydrogen
atom is to atomic physics. In most practical situations they
both become the first example where any respective new
formalism, description or theory is brought to be checked out.
Actually, since the very moment it was discovered one finds it
difficult to encounter a paper on cosmology in which no mention
is made of de Sitter universe. Soon after its publication [1],
Einstein sent a series of letters to de Sitter [2] in which he
did not express entire approval to the de Sitter cosmology. Two
main criticisms were raised by Einstein. Firstly, since it could
only apply without any 'world material', that is stars, he found
the de Sitter metric to make no physical meaning, and secondly,
Einstein argued that whereas one can always put a rigid circular
hoop into de Sitter universe at any sufficiently early, though
nonzero time, there is no place in it at time $t=0$. While the
first criticism leaves de Sitter universe only as a good
approximation for sufficiently tiny matter density (which
actually is the real case), the second one is nothing but an
expression of the Einstein disappeal to the existence of the big
bang, an attitude which evidently overcomes de Sitter scenario.

It is rather ironic however that, in spite of the Einstein
reservations, de Sitter obtained his cosmological solution by
just introducing the cosmological term first considered by
Einstein himself in the general-relativity field equations for a
spherically symmetric spacetime without any matter, that is by
considering the simplest possible scanario meeting the
essentials of Einstein ideas. The key differential point between
de Sitter and Einstein [3] universes is that in the latter a
given matter density was explicitly included, setting a
preferred value of the cosmological constant so that the
resulting solution represented a static universe. On the
contrary, in de Sitter space the value of the cosmological
constant is completely arbitrary. In the absence of any matter,
the Einstein equations can be written as
\begin{equation}
R^{\mu\nu}=\Lambda g^{\mu\nu} ,
\end{equation}
with $\Lambda$ the cosmological constant. Assuming then
staticness for a general spherically symmetric metric
\begin{equation}
ds^2=-e^{A}dt^2+e^{B}dr^2+r^2 d\Omega_2^2,
\end{equation}
with $d\Omega_2^2=d\theta^2+\sin^2 \theta d\phi^2$ the metric on
the unit three-sphere, we obtain $A=-B$ and hence the
Schwarschild-de Sitter solution [4]
\begin{equation}
e^{A}=e^{-B}=1-\frac{\Lambda r^2}{3}-\frac{2m}{r} ,
\end{equation}
where the mass $m$ comes about as an integration constant. de
Sitter's static solution arises when we set $m=0$ in Eq. (1.3),
i.e. in the pure vacuum case.

From a slightly more technical point of view, the de Sitter
space can be indentified [5] as a maximally symmetric space of
constant positive curvature (positive Ricci scalar) which is a
solution to the Einstein equations (1.1) with a positive
cosmological constant $\Lambda>0$. It can be visualized as a
five-hyperboloid [6],
\begin{equation}
-v^2+w^2+x^2+y^2+z^2=\frac{3}{\Lambda} ,
\end{equation}
embedded in $E^5$, with the most general metric of the de Sitter
space being then that which is induced in this embedding, i.e.
\begin{equation}
ds^2=-dv^2+dw^2+dx^2+dy^2+dz^2 .
\end{equation}
This has topology $R\times S^4$, invariance group $SO(4,1)$, and
shows ten Killings vectors (four boosts and six rotations).

What we have hitherto described corresponds to a de Sitter space
with simply connected topology. Recently, Gott and Li [7] have
also considered the important case of a multiply connected de
Sitter space. These authors imposed the identification symmetry
of the universal covering of Misner space [5] to the
five-hyperboloid visualizing the de Sitter space, so inducing a
boost transformation in the region $w>|v|$ covered by the static
metric of this space that makes this region multiply connected.
In this way, the event horizon of the de Sitter space becomes a
(Cauchy) chronology horizon which defines the onset of an
interior nonchronal region which supports closed timelike curves
(CTC's). The multiply connected de Sitter space can be used to
implement new boundary conditions for the universe according to
which, rather than being created from nothing, the universe
created itself [7].

In this paper we argue that the static de Sitter space for all
topologies actually corresponds to rather restricted solutions
of the Einstein field equations for a spherically symmetric
empty space which becomes curved only by the action of quantum
vacuum effects. We show that there exist more general solutions
to this problem, defined by two cosmological parameters, namely
the usual cosmological constant, and a new constant vacuum term
could could be associated to vacuum effects whose backreaction
over the spacetime geometry led to higher-derivative terms in
the gravitational action.

The rest of the paper can be outlined as follows. In Sec. II we
discuss the physical arguments that lead to obtaining a
two-parameter, generalized de Sitter solution, deriving the
precise form of such a new solution. We investigate then the
nature of the event horizons and the global structure of the new
solution, and maximally extend the metric beyond the apparent
horizons, both in the simply connected and multiply connected
cases. The thermal properties associated with the existence of
these event horizons are studied in Sec. III by using the usual
Euclidean procedure, and a new method based on replacing the
Kruskal maximal extension by the six-dimensional embedding
Minkowskian space as the background spacetime where we allow the
propagation of a massless scalar field. In Sec. IV we
investigate the effects that different types of Ginsparg-Perry
perturbations [8] have in the generalized Schwarzschild-de
Sitter space, concluding that this space is stable to the action
of such perturbations in first order. Finally, we briefly
summarize and conclude in Sec. V.

\section{\bf A two-parameter vacuum solution}
\setcounter{equation}{0}

We start this section by noticing the sense in which the de
Sitter space cannot be considered as the most general solution
to spherically symmetric Einstein equations for the pure vacuum
case. As it was pointed out in the Introduction, de Sitter space
can be visualized by embedding it in the five-hyperboloid (1.4),
with embedding Minkowski metric (1.5). However, the general
embedding class for metrics of type (1.2), with
$A=-B=\ln[1+\rho(r)]$ and $\rho(r)$ a given function of the
radial coordinate $r$, has been shown by Stephani [9] to be two,
and therefore one should consider a six-dimensional Minkowski
space as the most general embedding for it. This would mean that
one of the six degrees of freedom of the embedding space should
become frozen in the case of the usual de Sitter metric, so
pointing toward some lack of generality for this metric. In
fact, let us take the function $\rho(r)$ to be given by
$\rho(r)=-H^2 r^2$, with $H=\sqrt{\Lambda/3}$, and write the
six-dimensional Minkowski metric as
\begin{equation}
ds^2=-dz_0^2 +dz_i^2 , \;\; i=1,...,5
\end{equation}
embedding the de Sitter manifold according to e.g. the following
coordinate transformations
\begin{equation}
z_0=k\sqrt{1-H^2 r^2}\sinh(t/k),\; z_1=k\sqrt{1-H^2
r^2}\cosh(t/k)
\end{equation}
\begin{equation}
z_2=f(r)
\end{equation}
\begin{equation}
z_3=r\cos\phi\sin\theta,\; z_4=r\sin\phi\sin\theta, \;
z_5=r\cos\theta ,
\end{equation}
where $k$ is an arbitrary, dimensional parameter to be fixed
later, and the function $f(r)$, defining the new coordinate
$z_2$, is given by [10]
\begin{equation}
\left(\frac{df}{dr}\right)^{2}=\left(f'\right)^{2}=
-\left[\frac{k^2 \left(d\rho/dr\right)^2
+4\rho}{4(1+\rho)}\right] .
\end{equation}
Taking for the arbitrary parameter $k$ the inverse of the
surface gravity of de Sitter space, $k=\kappa_c^{-1}=H^{-1}$, we
get then from Eq. (2.5) that the function $f(r)$ becomes a
constant, $f_0$, for de Sitter space, and hence it follows that
the coordinate $z_2$ no longer is a degree of freedom for this
space. Therefore, though the de Sitter static metric can still
be visualized as a six-hyperboloid, this is reducible to a
five-hyperboloid defined for a re-scaled cosmological constant
$\Lambda_R=1/\left(\Lambda^{-1}-f_0^2/3\right)$, and the
embedding Minkowski metric becomes that of a five-dimensional
space. In what follows we interpret this result as being an
indication that the known de Sitter solution is not the most
general solution that corresponds to general relativity with no
matter for consistent nonzero quantum-mechanical vacuum
fluctuations, for the following reason. Einstein equations were
initially thought of as satisfying the requirement that they
should permit flat spacetime as a particular solution in the
absence of matter. By introducing the cosmological constant,
$\Lambda$, Einstein dropped [3] this requirement out and got a
slightly more general set of equations which do not allow the
existence of globally flat spacetime in the absence of matter,
for then these equations become identical to Eq. (1.1).

Motivated by the facts that (i) any cosmological constant terms
should all be associated to quantum fluctuations of vacuum [11],
(ii) the gravitational Lagrangian must contain curvature squared
corrections whenever it is considered quantum-mechanically [12],
and (iii) the physical dimensions of the cosmological constant
terms should be that of the non-constant part of the geometric
terms containing the curvature in the gravitational action, we
now go a step further by introducing in spherically symmetric
spaces another vacuum term which we assume to depend on the
radial coordinate, $r$, as $\tilde{Q}r^2$, where the constant
$\tilde{Q}$ can, in principle, be positive, negative or zero.
Like the curvature squared terms $R^2$, $R_{\mu\nu}R^{\mu\nu}$,
etc., in higher-derivative gravity theories, the new constant
$\tilde{Q}$ ought to have the physical dimensions of a space
curvature squared, namely (length)$^{-4}$. Disregarding the
curvature squared terms in the left-hand-side of the so modified
field equations, we obtain then
\begin{equation}
R^{\mu\nu}=g^{\mu\nu}\left(\Lambda+\tilde{Q}r^2\right) .
\end{equation}
Strictly speaking, one had to have considered field equations
associated with higher-derivative theories with $R^2$ terms.
However, in order to simplify our formulation, we restrict here
ourselves to work in a semiclassical approximation where all
quantum-mechanical effects are only taken into account in the
right-hand-side of the field equations.

Solutions obtained from Eq. (2.6) will then depend on two
parameters, the cosmological constant $\Lambda$ and the new
parameter $\tilde{Q}$. The physical meaning of the cosmological
constant is explicitly obtained in the quantum field theory
context as being the (homogeneous and isotropic) ground state
expectation value of some vacuum stress tensor. As to the
physical interpretation of $\tilde{Q}$, one may claim it to have
an analogous origin as for $\Lambda$. Thus, $\tilde{Q}$ ought to
be quantum-mechanically associated with some additional
component of the vacuum stress tensor which would allow its
value to be nonvanishing. On the other hand, given the explicit
radial dependence assumed for the additional term which is
displayed in Eq. (2.6), and the existence of a time-like
singularity as one approaches $r=+\infty$ (see Fig. 1), one
could expect that a nonvanishing $\tilde{Q}$ would break
homogeneity of the spatial sections at very large scales. This
seems to raise the interesting possibility that corrections
coming from $\tilde{Q}$ might be considered as fluctuations
around a homogeneous space, which eventually could be regarded
as the seeds required for the formation of galaxies. Although in
order to fix the precise form and physical meaning of the
relation between the solutions containing the parameter
$\tilde{Q}$ and quantum vacuum fluctuations would require
further detailed studies, it is still at this stage possible to
conjecture that the generalized de Sitter spaces must describe
additional quantum fluctuations which can be associated with
virtual charged particles located at $r=0$. Note furthermore
that the way in which $\tilde{Q}$ will enter the relevant
solutions (see e.g. Eq. (2.7)) remainds one of the electric
charge in Reissner-Nordstrom black hole solution.

Assuming then staticness, for the general spherically symmetric
metric (1.2), we shall again obtain from the field equations
(2.6) that $A=-B$, and from the equation for the spherical angle
$\theta$,
\begin{equation}
e^{A}=e^{-B}= 1-\frac{\Lambda
r^2}{3}-\frac{\tilde{Q}r^4}{5}-\frac{2m}{r} ,
\end{equation}
with $m$ an integration constant playing the role of an existing
mass. Solution (2.7) describes a generalized Schwarzschild-de
Sitter spacetime.

In what follows of this section and in the next one we consider
the case $m=0$. Then, depending on the sign of the involved
constants, we have distinct kinds of solutions. Setting
$H=\sqrt{|\Lambda|/3}$ and $Q^4=|\tilde{Q}|/5$, if $H^2>0$ and
$\tilde{Q}>0$, the resulting solution shows an event horizon at
\[r_H^2=\frac{\sqrt{H^4+4Q^4}-H^2}{2Q^4} ;\]
if $H^2>0$ and $\tilde{Q}<0$, then there will be two horizons at
\begin{equation}
r_{\pm}^2=\frac{H^2\pm\sqrt{H^4-4Q^4}}{2Q^4} ,
\end{equation}
with $H^2\geq 2Q^2$. Finally when $H^2<0$ and $\tilde{Q}>0$
(generalized anti-de Sitter solution), we obtain also an event
horizon at
\[r_H^2=\frac{H^2+\sqrt{H^4+4Q^4}}{2Q^4}.  \]
In this paper, we will only consider the second of these
particular solutions, i.e.
\[ds^2=-\left(1-H^2
r^2+Q^4 r^4\right)dt^2\]
\begin{equation}
+\left(1-H^2 r^2+Q^4
r^4\right)^{-1}dr^2+r^2 d\Omega_2^2 .
\end{equation}
This metric can be embedded in the six-dimensional Minkowski
space with metric (2.1) using the coordinate transformations
\[z_0=k\sqrt{1-H^2 r^2+Q^4 r^4}\sinh(t/k),\]
\begin{equation}
z_1=k\sqrt{1-H^2 r^2+Q^4 r^4 }\cosh(t/k)
\end{equation}
\begin{equation}
z_2=f(r)
\end{equation}
\begin{equation}
z_3=r\cos\phi\sin\theta,\; z_4=r\sin\phi\sin\theta, \;
z_5=r\cos\theta ,
\end{equation}
where the function $f(r)$ again satisfies condition (2.5) which
is singular at $r_{\pm}$. Taking for the constant $k$ the
inverse of the surface gravity which now becomes
\begin{equation}
\kappa_c=\frac{r_+ -r_-}{r_+^2}=
\frac{2Q^2\sqrt{H^2-2Q^2}}{H^2 +\sqrt{H^4-4Q^4}} ,
\end{equation}
we get by using the equalities $Q^2=1/r_+ r_-$,
$H^2=(r_+^2+r_-^2)/r_+^2 r_-^2$, and $\sqrt{H^4-4Q^4}=(r_+^2
-r_-^2)/r_+^2 r_-^2$,
\[(f')^2=
-\left[\frac{r_+ r_-^3\left(r_+^2
+r_-^2\right)\left(\frac{r_+^3}{r_-^3}
-\frac{r_-}{r_+}+2\right)r^2}{r_+^2 r_-^4\left(r_+ -r_-\right)^2
g_{tt}}\right.\]
\begin{equation}
+\left.\frac{r_-^2\left[\left(r_+ -r_-\right)^2
-2r_+^2\left(\frac{r_+^2}{r_-^2}+1\right)\right]r^4 +4r_+^2
r^6}{r_+^2 r_-^4\left(r_+ -r_-\right)^2 g_{tt}}\right] ,
\end{equation}
which only vanishes when $Q$ approaches zero. It appears that
all of the degrees of freedom of the six-hyperboloid are now
being used and that, therefore, metric (2.9) is more general
than the usual de Sitter metric in order to describe a
spherically symmetric empty space which is subjected to quantum
vacuum fluctuations.

The function $(f')^2$ is positive and nonsingular for $r\leq
r_-$, i.e. in the region of the generalized de Sitter space
which can be thought of as a four-dimensional membrane immersed
in a six-dimensional flat space, with the parametric equations
given by
\begin{equation}
z_1^2-z_0^2=\frac{Q^4 r_+^4}{H^2-2Q^2}\left(1-H^2 r^2+Q^4
r^4\right)
\end{equation}
\begin{equation}
z_2=f(r)
\end{equation}
\begin{equation}
z_3^2+z_4^2+z_5^2=r^2 .
\end{equation}
The chosen solution $H^2 >0$, $\tilde{Q}<0$ is valid only for
$H^2\geq 2Q^2$. As $H^2$ approaches $2Q^2$ one obtains an {\it
extreme} generalized de Sitter space, a case which we shall
briefly discuss later on.

The transformations (2.10) are not unique and cover only the
region $z_1>|z_0|$ of the Minkowski space. There could actually
be four distinct transformations that together cover completely
the Minkowski space, namely
\[z_0=\pm k\sqrt{1-H^2 r^2+Q^4 r^4}\sinh(t/k) ,\]
\[z_1=\pm k\sqrt{1-H^2 r^2+Q^4 r^4}\cosh(t/k) \]
plus another two with the coordinates $z_1$ and $z_0$
interchanged ($z_1\leftrightarrow z_0$). This is exactly the
same situation as that we are going to find when extending from
metric (2.9) to its associated Kruskal metric [10]. In fact,
metric (2.9) is geodesically incomplete because of the presence
of the event horizons at $r_{\pm}$, as given by Eq. (2.8). This
geodesic incompleteness can be avoided by using the Kruskal
technique [5,13], and therefore we first introduce the
"tortoise" coordinate
\[r^{*}=\int\frac{dr}{1-H^2 r^2+Q^4 r^4}\]
\begin{equation}
=\frac{r_+^2 r_-^2}{2\left(r_+^2
-r_-^2\right)}\left[\frac{1}{r_+}\ln\left(\frac{r_+ -r}{r_+
+r}\right)-\frac{1}{r_-}\ln\left(\frac{r_- -r}{r_-
+r}\right)\right] ,
\end{equation}
and then the advanced and retarded coordinates
\[V=t+r^{*} ,\;\;\; W=t-r^{*} ,\]
so that the metric becomes
\begin{equation}
ds^2=-\left(1-H^2 r^2+Q^4 r^4\right)dVdW+r^2 d\Omega_2^2 ,
\end{equation}
with the radial coordinate, $r$, defined by
\[r^{*}=\frac{1}{2}(V-W) .\]
In the case that $Q^2<H^2/2$, we introduce now the new
coordinates $V'$ and $W'$ by means of the definitions
\begin{equation}
V'=\tan^{-1}\left[\exp(\kappa_c V)\right] ,\;\;
W'=\tan^{-1}\left[-\exp(\kappa_c W)\right] ,
\end{equation}
and hence we obtain the Kruskal metric for a generalized de
Sitter space with negative constant $\tilde{Q}$, that is
\[ds^2=\]
\begin{equation}
-\left(1-H^2 r^2+Q^4 r^4\right)\frac{4dV'dW'}{\kappa_c^2
\sin(2V')\sin(2W')}+r^2 d\Omega_2^2 ,
\end{equation}
where $\kappa_c$ is the surface gravity defined in Eq. (2.13),
and $r$ is defined implicitly by the relation:
\begin{equation}
\tan V'\tan W'= -\left[\left(\frac{r_+ -r}{r_+
+r}\right)^{\frac{r_-}{r_+}}\left(\frac{r_- +r}{r_-
-r}\right)\right]^{\frac{r_-}{r_+ +r_-}} .
\end{equation}
The maximal extension is therefore obtained taking metric (2.21)
as the metric of the largest manifold which metrics given only
either in terms of $V$ or in terms of $W$ can be isometrically
embedded. There will be then a maximal manifold on which metric
(2.21) is $C^2$. One can also use the conformal treatment of the
infinities to produce the Penrose diagram corresponding to the
generalized de Sitter space we are dealing with. It is depicted
in Fig. 1, where the regions labelled $I$ describe the interval
$0<r<r_-$, the regions labelled $II$ describe the interval $r_-
<r<r_+$, and the regions labelled $III$ describe the spacetime
within the exterior interval $r_+ <r<\infty$. This diagram is
formed up as an infinite chain of asymptotically flat
($r\rightarrow 0$) regions $I$ which are connected to one
another by means of regions $II$ and $III$, and each region
$III$ is bounded by a timelike asymptotically flat infinite
singularity at $r\rightarrow\infty$.

So far, we have confined ourselves to the case of a generalized
de Sitter space with $\tilde{Q}<0$ for a simply connected
topology. One can also consider this spacetime with multiply
connected topology on a given restricted region of it. This can
be accomplished by imposing the symmetry of the Misner space [5]
to the visualizing six-dimensional hyperboloid that embeds the
generalized de Sitter space, that is by letting the Minkowski
coordinates of that hyperboloid satisfy the identification
property
\[\left(z_0,z_1,z_2,z_3,z_4,z_5\right)\leftrightarrow \]
\[\left(z_0\cosh(nb)+z_1\sinh(nb),z_0\sinh(nb)+z_1\cosh(nb),\right.\]
\begin{equation}
\left.z_2,z_3,z_4,z_5
\right) ,
\end{equation}
where $b$ is a dimensionless arbitrary period and $n$ is any
integer number, $n=0,1,2,...,\infty$. The boost transformation
in the $(z_0,z_1)$ plane implied by identifications (2.23) will
induce a boost transformation in the generalized de Sitter
space. Hence, since the boost group in the generalized de Sitter
space must be a subgroup of the generalized de Sitter group, the
static metric (2.9) should also be invariant under the symmetry
(2.23). One can readily see that this symmetry is in fact
satisfied in the generalized de Sitter region covered by metric
(2.9) and defined by $z_1 >|z_0|$, where there are CTC's, with
the boundaries at $z_1=\pm z_0$ and a given value of
$\sum_{i=2}^{5}z_i^2$ being the Cauchy horizons that limit the
onset of the nonchronal region from the causal exterior; that is
the chronology horizons for the multiply connected generalized
de Sitter space [7]. As applied to this space, the
identifications (2.23) become
\begin{equation}
t\leftrightarrow t+nb\kappa_c=
t+\frac{nb\left(H^2+\sqrt{H^4-4Q^4}\right)}{2Q^2\sqrt{H^2-2Q^2}},
\end{equation}
which applies only to the region defined by $r\leq r_-$. In the
case of an extreme generalized de Sitter space, where
$Q^2=H^2/2$, identifications (2.24) reduce to $t\leftrightarrow
t+\infty$, a result which corresponds to the fact that the
temperature of the extreme generalized de Sitter space is zero.

\section{\bf Thermal radiation in generalized de Sitter space}
\setcounter{equation}{0}

In order to study the thermal properties of the generalized de
Sitter space with $\tilde{Q}<0$, which are to be expected
because of the existence of event horizons in this space, we
shall follow two distinct procedures. We shall first use the
conventional Euclidean method [14], and then a new method based
on calculating the Fourier transform of the Whitman function
that corresponds to the propagation of a massless scalar field
in the six-dimensional Minkowskian embedding space [10,15].

We start with the Kruskal extension of the static metric which
does not contain the geodesic incompleteness at the event
horizons. In order for this metric to be definite positive we
have to analytically continue the time coordinate according to
the rotation $t=i\tau$. To see how this can be accomplished, we
introduce new coordinates in the metric (2.21), such that
\begin{equation}
\tan V'=R+S ,\;\;\; \tan W'=R-S .
\end{equation}
Then, the metric (2.21) becomes
\begin{equation}
ds^2=\frac{\left(1-H^2 r^2+Q^4
r^4\right)\left(dR^2-dS^2\right)}{\kappa_c^2\left(R^2-S^2\right)}
+r^2d\Omega_2^2 ,
\end{equation}
where $r$ is defined by
\begin{equation}
S^2-R^2= \left[\left(\frac{r_+ -r}{r_+
+r}\right)^{\frac{r_-}{r_+}}\left(\frac{r_- +r}{r_-
-r}\right)\right]^{\frac{r_-}{r_+ +r_-}} ,
\end{equation}
so that the surface $r=0$ is defined by $S^2 -R^2=1$, and the
time $t$ results from
\begin{equation}
\frac{R-S}{R+S}=-\exp\left(-2\kappa_c t\right) .
\end{equation}
In the region $r\leq r_-$, metric (3.2) becomes definite
positive, provided we perform the continuation
\begin{equation}
S=i\xi ,\;\;\; t=i\tau .
\end{equation}
Then, from Eq. (3.4) we obtain
\begin{equation}
\xi +iR=-\sqrt{R^2+\xi^2}\exp\left(-i\kappa_c \tau\right) .
\end{equation}
We thus see that the Euclidean time $\tau$ becomes a periodic
variable, with a period $2\pi/\kappa_c$. Following then the
usual procedure we can deduce that the generalized de Sitter
space with $\tilde{Q}<0$ is characterized by a background
temperature given by
\begin{equation}
T=\frac{\kappa_c}{2\pi}=
\frac{Q^2\sqrt{H^2-2Q^2}}{\pi\left(H^2+\sqrt{H^4-4Q^4}\right)} ,
\end{equation}
where it can be noted that in the extreme case $Q^2=H^2/2$,
$T=0$, so that no thermal bath remains.

Although it reproduces the correct results, the Euclidean method
is rather mysterious and too much schematic. To see the physics
behind the thermal radiation process, we shall now follow a
different procedure which relies on performing a calculation in
the six-dimensional Minkowski space [10,15], which is thereby
taken to play the role of the maximum extension of the static
metric, instead of the Kruskal space on which our previous
discussion was made.

As related to the embedding of the generalized de Sitter space
with $\tilde{Q}<0$, the Rindler wedge restriction of the
six-dimensional Minkowski space requires the usual coordinate
re-definitions
\[z_0=\zeta\sinh(g\eta) ,\;\; z_1=\zeta\cosh(g\eta), \]
with all other $z_m$ remaining unchanged. Then,
\begin{equation}
ds^2=-g^2\zeta^2 d\eta^2+d\zeta^2+\sum_{m=2}^{5}dz_m^2 .
\end{equation}
The lines $\zeta=$constant, $z_m=$constant are the trajectories
of the uniformly accelerated observers with constant
acceleration $g$; i.e. those observers moving along the
hyperbolas $\zeta^2=z_1^2-z_0^2$. They will possess an event
horizon located at $\zeta=0$. The coordinates $\eta$, $\zeta$
cover only the quadrant $z_1>z_0$, such as the coordinates $t$,
$r$ do. One can pass from metric (3.8) to the static metric by
using
\begin{equation}
\eta=t ,\;\; \zeta=\frac{\sqrt{1-H^2 r^2+Q^4 r^4}}{\kappa_c} ,
\end{equation}
so as the relations for $z_2$, $z_3$, $z_4$ and $z_5$ given in
Sec. II, and the identification $g=\kappa_c$. Because $\zeta=$
constant represents a trajectory with constant proper
acceleration, we conclude that a static observer ($r$, $\theta$,
$\phi$ constants) in the interior generalized de Sitter static
geometry is a particular type of uniformly accelerated observer
(with acceleration proportional to $\kappa_c$) in the embedding
Minkowski space.

The power spectrum detected by an accelerated de Witt detector
which probes the quantum fluctuations of a scalar massless field
is given by [16]
\begin{equation}
F(\omega)\int_{-\infty}^{+\infty}D(\eta,\eta
')\exp(i\omega\bigtriangleup\eta)d(\bigtriangleup\eta) ,
\end{equation}
where $D(\eta,\eta ')$ is the Whitman function $D(z,z')$ [14]
evaluated along the world line of the detector,
$\bigtriangleup\eta=\eta-\eta '$. There is now the problem that,
although the world line of the detector is the same in Minkowski
and generalized de Sitter spaces, the scalar field is not
restricted to the two-dimensional space, but extends over the
full spacetime. In order to keep studying the behaviour of the
field close to the detector world line, we will make our
calculation in the geometric optics approximation where only
short wavelengths are taken into account [10]. Thus, in this
approximation, the Whitman function for the world line (2.10)
becomes
\[D(\eta,\eta ')=\]
\begin{equation}
 -\frac{\kappa_c^2}{16\pi^2\left(1-H^2
r^2+Q^4
r^4\right)\sinh^2\left[\kappa_c(\bigtriangleup\eta-i\epsilon)/2\right]}
.
\end{equation}
Choosing as the integration contour in the $\bigtriangleup\eta$
complex plane the strip $0\leq Im\kappa_c\bigtriangleup\eta\leq
2\pi$ and using the theorem of residues, we can evaluate the
Fourier integral (3.10) to give
\begin{equation}
F(\omega)\propto
\frac{1}{\left(1-H^2 r^2+Q^4
r^4\right)\left[\exp\left(\frac{2\pi\omega}{\kappa_c}\right)-1\right]}
,
\end{equation}
the proportionality factor being given by a certain product of
powers of the frequency $\omega$ and the trajectory of the
accelerated observer $\zeta$. In this approximation, the power
spectrum is thus Planckian with a temperature given by
expression (3.7). Making the same kind of calculation in the
full six-dimensional Minkowski space, without resorting to
geometric optics approximation, we finally obtain for the power
spectrum
\begin{equation}
F(\omega)\propto
\frac{1+\frac{\kappa_c^2
}{\omega^2}}{\exp\left(\frac{2\pi\omega}{\kappa_c}\right)-1} ,
\end{equation}
which is not Planckian and only becomes so in the geometric
optics approximation where $\omega\gg\kappa_c$. The
proportionality factor in (3.13) also depends on a product of
given powers of frequency $\omega$ and observer's trajectory
$\zeta$. It is worth noting the dependence of $F(\omega)$ on
$\zeta$ which reflects the feature that the horizons in our
spacetime are observer-dependent, such as it happens in de
Sitter space.

\section{\bf Generalized Schwarzschild-de Sitter space}
\setcounter{equation}{0}

In this section we briefly discuss the different possible
first-order perturbations of the Ginsparg-Perry type [8] that
can occur in both the singly degenerate and doubly degenerate
cases corresponding to the generalized Schwarzschild-de Sitter
space. Let us recall that the metric of this space can be
defined by
\begin{equation}
e^{A}=V(r)=1-H^2 r^2+Q^4 r^4 - \frac{2m}{r} .
\end{equation}
In order to determine the positions of the horizons in the
extreme (degenerate) case, we first obtain the extremals of the
function $rV(r)$, i.e. we solve for $d\left[rV(r)\right]/dr=0$,
which gives for $Q^2=H^2/2$
\begin{equation}
\left(r_{\pm}^0\right)^2=\frac{2(3\pm 2)}{5H^2} .
\end{equation}
The horizon at $r_+^0$ corresponds to the singly degenerate case
(the two cosmological horizons coincide). From the horizon
condition $V(r_+^0)=0$ we obtain then $m=0$, i.e. there is no
black hole in this case, and $r_+^0$ exactly is the degenerate
radius of the two cosmological horizons of the generalized de
Sitter space studied in the precedent sections. The horizon at
$r_-^0$ can be associated with the doubly degenerate case (the
degenerate cosmological horizon also coincides with the radius
of the black hole). Replacing $r_-^0$ into $V(r)=0$, we get then
\begin{equation}
m=m_0=\frac{8}{5}r_-^0 .
\end{equation}

We consider now the distinct first-order perturbations of the
type first considered by Ginsparg and Perry [8]for the above two
degenerate cases. Let us first study perturbations of the doubly
degenerate case where we first introduce the ansatz
\[t=\frac{\psi}{2r_-^0 H^2\delta} \]
\begin{equation}
r=r_-^0\left(1-\delta\cos\chi+\frac{1}{13}\delta^2\right)
\end{equation}
\[Q^4=H^4\left(\frac{1}{4}-\frac{40}{13}\delta^2\right) ,\]
with $m_0$ unperturbed, and $0<\delta\ll 1$. In this case, we
have
\begin{equation}
V(\chi)=\frac{4}{5}\delta^2\sin^2\chi ,
\end{equation}
so that at first order in $\delta$,
\begin{equation}
ds^2=
\frac{1}{2H^2}\left(d\chi^2-\sin^2\chi d\psi^2\right)
+\left(r_-^0\right)^2\left(1+2\delta\cos\chi\right)d\Omega_2^2 .
\end{equation}
As fixed by $V(\chi)=0$, the horizons will appear at $\chi=0$
and $\chi=\pi$, which correspond to the perturbed cosmological
horizons at $r_+ =r_-^0(1+\delta)$ and $r_- =r_-^0(1-\delta)$,
respectively. The radius of the black hole keeps being located
at $r_b=r_-^0$ and it no longer is associated to an event
horizon.

In order to evaluate the thermal properties of the so-perturbed
spacetime, we calculate the surface gravity, $\kappa$, in each
case, determining then the associated temperature by means of
the simple expression $T=\kappa/2\pi$. In order to compute
$\kappa$, we shall follow Bousso and Hawking [17], i.e. we set
\begin{equation}
\kappa=2\pi\frac{\gamma_{\psi}}{\psi_h^{id}} ,
\end{equation}
where
\begin{equation}
\gamma_{\psi}=H^2 r_-^0
\end{equation}
and
\begin{equation}
\psi_h^{id}=
\left. 2\pi\sqrt{g_{\chi\chi}}\right|_{\chi=\chi_h}
\left.\left(\frac{\partial}{\partial\chi}
\sqrt{g_{\psi\psi}}\right|_{\chi=\chi_h}\right)^{-1} .
\end{equation}
We then obtain
\[T_+ =\frac{H^2 r_-^0}{2\pi} ,\;\; T_- =-\frac{H^2
r_-^0}{2\pi},\;\; T_b=0 ;\] i.e. the black hole with
intermediate size would not radiate, but there is a radiation
process transfering energy from the outer to the inner horizon
of the generalized de Sitter space, until the double degeneracy
is restored. Thus, the system appears to be stable to these
perturbations.

We next consider perturbations of the form
\[r=r_-^0\left(1-\delta\cos\chi-\delta^2\right) \]
\begin{equation}
m=m_0\left(1+\frac{11Hr_-^0\delta^2}{8}\right)
\end{equation}
\[Q^4=H^4\left(\frac{1}{4}-2\delta^2\right) ,\]
with the time coordinate $\psi$ as defined by the first of Eqs.
(4.4). Hence, one again obtains Eqs. (4.5) and (4.6) for
$V(\chi)$ and the perturbed metric, respectively. Thus, the
horizons at $\chi=0,\pi$ will correspond to the respective radii
$r_h(\chi=0)=r_-^0(1+\delta)$ and
$r_h(\chi=\pi)=r_-^0(1-\delta)$. This may describe two distinct
situations. Either (a) $r_h(\chi=0)=r_b$ (radius of an
evaporating black hole) and $r_h(\chi=\pi)=r_{-b}$ (radius of an
anti-evaporating balck hole [18]), with the degeneracy of the
cosmological horizons preserved at $r_{\pm}=r_-^0$, or (b)
$r_h(\chi)=r_+=r_b$ and $r_h(\chi=\pi)=r_-=r_{-b}$, i.e. the
degeneracy of the cosmological horizons is broken, but that
between the horizons of the evaporating (anti-evaporating) black
hole and the cosmological outer (inner) horizons is preserved.
In case (a) the temperatures of the black holes, as calculated
by using Eqs. (4.7) - (4.9), are $T_{\pm b}=\pm H^2 r_-^0/2\pi$,
and those of the degenerate cosmological horizons are $T_ {\pm}=0$.
In case (b), we obtain instead, $T_{\pm b}=T_{\pm}=\pm H^2
r_-^0/2\pi$. It follows that in any of the two cases, our
spacetime is stable to the perturbations specified by Eqs.
(4.10).

Another kind of possible perturbations in the doubly degenerate
case can finally be considered. It would describe perturbations
which always leave the degeneracy between the cosmological
horizons unchanged, and is defined by the same time perturbation
as in Eqs. (4.4) and (4.10), and
\[r=r_-^0\left(1+\delta\cos\chi\right) \]
\begin{equation}
m=m_0\left(1-\frac{5}{4}\delta^2\right)
\end{equation}
\[Q^4=\frac{1}{4}H^4 .\]
These perturbations again lead to Eqs. (4.5) and (4.6) for
$V(\chi)$ and the perturbed metric. In this case, the
temperatures that result are: $T_{\pm}=0$ and $T_{\pm b}=\pm H^2
r_-^0/2\pi$, so that the spacetime is once again stable to the
perturbations.

Let us consider in what follows of this section the case in
which the horizon is at $r=r_+^0$, with $m=0$. It describes a
singly degenerate generalized de Sitter space with negative
$Q^4$ and no black hole being initially present. We introduce
then the perturbative parameter $\delta$ by means of the
equation:
\begin{equation}
Q^4 =H^4\left(\frac{1}{4}-\frac{2}{5}\delta^2\right) ,\;\;
0\leq\delta\ll 1 .
\end{equation}
The singly degenerate case corresponds to the limit
$\delta\rightarrow 0$. The new time and radial coordinates,
$\psi$ and $\chi$, will now be given by
\begin{equation}
t=\frac{\psi}{2r_+^0 H^2\delta} ,\;\;
r=r_+^0\left(1+\delta\cos\chi\right) .
\end{equation}
We then let the perturbation induce creation (or annihilation)
of an extremely tiny black hole with mass
\begin{equation}
m=\frac{6r_+^0\delta^2}{5} .
\end{equation}
The inner and outer horizons of the generalized de Sitter space
lie, respectively, at $\chi=\pi$ and $\chi=0$, while the balck
hole horizon will be at $2m$. To first order in $\delta$, from
the above transformations, we obtain the new metric
\[ds^2=\frac{1}{2H^2}\left[(1+2\delta\cos\chi)\sin^2\chi
d\psi^2\right. \]
\begin{equation}
\left.-(1-2\delta\cos\chi)d\chi^2\right]
+\frac{2}{H^2}(1+2\delta\cos\chi)d\Omega_2^2 .
\end{equation}
We notice that, since $V(r)=1-H^2 r^2+Q^4 r^4$ becomes negative
when it is perturbed, i.e.
\begin{equation}
V(\chi)=-4\delta^2\left(1+2\delta\cos\chi\right)\sin^2\chi,
\end{equation}
the coordinate $\psi$ becomes spacelike and the radial
coordinate becomes timelike. In these coordinates, the topology
of the sections at constant $\psi$ is $S^1\times S^2$, and this
means creation of a handle. For these solutions, the radius of
the two-spheres, $r$, varies along the $S^1$ coordinate $\chi$,
with the minimal (maximal) two-sphere corresponding to the inner
(outer) cosmological horizon, and the ergoregion between the
cosmological horizons nesting a little black hole of radius $2m$
much smaller than the size of that ergoregion $r_+
-r_-=2r_+^0\delta$. At the degenerate case, the metric (4.14)
becomes
\begin{equation}
ds^2=\frac{1}{2H^2}\left(\sin^2\chi d\psi^2-d\chi^2\right)
+\frac{2}{H^2}d\Omega_2^2 .
\end{equation}
This metric is not the same as the Nairai solution. One still
can recover a metric with the conventional signature by rotating
$\chi=i\sigma$, so that the line element (4.14) becomes
\[ds^2=-\frac{1}{2H^2}\left[(1+2\delta\cosh\sigma)\sinh^2\sigma
d\psi^2\right. \]
\begin{equation}
\left.-(1-2\delta\cosh\sigma)d\sigma^2\right]
+\frac{2}{H^2}(1+2\delta\cosh\sigma)d\Omega_2^2 ,
\end{equation}
and for the degenerate case
\begin{equation}
ds^2=-\frac{1}{2H^2}\left(\sinh^2\sigma d\psi^2-d\sigma^2\right)
+\frac{2}{H^2}d\Omega_2^2 .
\end{equation}
Metric (4.18) would represent a wormhole with the throat at the
outer horizon. If we Wick rotate the time variable, that is
$\psi=i\varphi$, we finally obtain a metric with Euclidean
signature. In the degenerate case, this will produce two round
two-spheres of different radius.

The interpretation of these perturbations becomes again twofold.
One can regard them as representing a breaking of the degeneracy
between the cosmological horizons at $r_{\pm}$, in which case
\[r_h(\chi=0)=r_+^0(1+\delta)=r_+=r_b \]
\[r_h(\chi=\pi)=r_+^0(1-\delta)=r_-=r_{-b} .\]
On the other hand, we can also look at these perturbations as
preserving that degeneracy, so that $r_h(\chi=0)=r_b$,
$r_h(\chi=\pi)=r_{-b}$, and $r_{\pm}=r_+^0$ (as evaluated at
$\chi=\pi/2$).

The calculation of the surface gravity for these perturbations
follows the same line as for the doubly degenerate cases, unless
for the parameter $\gamma_{\psi}$ which should now be given by
\begin{equation}
\gamma_{\psi}=H^2 r_+^0 ,
\end{equation}
instead of Eq. (4.8). We then obtain for the temperatures
corresponding to the two horizons $r_h$
\begin{equation}
T(\chi=0)=\frac{H^2 r_+^0(1+2\delta)}{2\pi}
\end{equation}
\begin{equation}
T(\chi=\pi)=-\frac{H^2 r_+^0(1-2\delta)}{2\pi} ,
\end{equation}
and for $\chi=\pi/2$,
\begin{equation}
T\left(\chi=\frac{\pi}{2}\right)=-\frac{H^2 r_+^0\delta}{2\pi} .
\end{equation}
When the degeneracy between the cosmological horizons is broken
by the perturbations, then $T_b=T_+=T(\chi=0)$,
$T_{-b}=T_-=T(\chi=\pi)$, while the temperature (4.22) measures
a thermal process by which the vacuum itself anti-radiates. In
the case for which the above degeneracy is preserved, we obtain
instead, $T_b=T(\chi=0)$, $T_{-b}=T(\chi=\pi)$, and
$T_{\pm}=T(\chi=\pi/2)$. All of these thermal processes lead to a
final state in which there is no black hole and the generalized
de Sitter space becomes finally again degenerate, so also these
perturbations are stable.

\section{\bf Summary and conclusions}
\setcounter{equation}{0}

In this paper we argue that the most general solution to the
general-relativity field equations that describe an empty
spherically symmetric spacetime must contain two constant
parameters in order to account for the quantum vacuum effects.
According to the possible combinations of the signs of such
parameters, the general solution splits into three essentially
different static spacetimes, two describing generalized de
Sitter spaces and one describing a generalized anti-de Sitter
space. In this paper, we have confined ourselves to the study of
the generalized de Sitter case (positive cosmological constant)
with a negative additional constant whose physical dimensions
are the same as those of a space curvature squared. Such a
solution possesses two event horizons and can be visualized as a
six-hyperboloid with embedding Minkowskian metric. It is
maximally extensible by defining Kruskal like coordinates, and
corresponds to a spacetime which can be made multiply connected
inside the inner event horizon by imposing the symmetry of the
Misner space.

Since the additional vacuum constant term in the right-hand-side
of the Einstein equations ought to correspond to an existing
curvature squared terms in the left-hand-side of such equations,
it could be thought that, as corresponding to all other
perturbation terms in the gravitational Lagrangian of a Lovelock
theory [19], one could have an arbitrary number of additional
constants characterizing quantum fluctuations of vacuum. From
this point of view, the solution studied in this paper might be
regarded as merely being nothing but the second-order
semiclassical approximation from the most general full solution.

The thermal properties of the spacetime dealt with in this work
have been studied by using both the conventional Euclidean
procedure, and a novel method based on considering the
propagation of a massless scaler field in its six-dimensional
embedding space, where the thermal spectrum is computed as the
Fourier transform of the Whitman equation. We obtain that an
observer moving along the origin of the radial coordinates will
detect a background thermal bath at a temperature which vanishes
only as one approaches the extreme (degenerate) case.

We have finally studied the stability of the generalized
Schwarzschild-de Sitter spacetime to the allowed Ginparg-Perry
perturbations in first-order approximation. It has been obtained
that this spacetime perdures to all types of the considered
perturbations, and that in the case that no balck hole was
initially present, processes are allowed in which black hole
pairs are created and then evaporated away.

There are two ways along which the present work could be
continued. On the one hand, it appears appropriate to
investigate the cosmological effects that the additional vacuum
term may have in cosmological models. One should then consider
the global Friedman-Robertson-Walker metric that can be
associated with the static metric studied in this work. On the
other hand, one would also investigate the effects that can be
expected from inserting a positive vacuum additional term in the
properties of the anti-de Sitter space. We have already obtained
that in this case, the generalized anti-de Sitter space has an
event horizon and, therefore, one should expect that, contrary
to what happens in the usual anti-de Sitter space, this space
has a precise temperature, and that its properties are similar
to those of the generalized de Sitter space with positive $Q^4$.

\acknowledgements
This research was supported by DGICYT under Research Project No.
PB97-1218.

\vspace{2cm}

\begin{center}
{\bf Legend for Figure}
\end{center}

\noindent {\bf Fig. 1}. The global structure of the generalized
de Sitter space with negative $\tilde{Q}$.

\end{document}